\documentclass[nofootinbib, aps,twocolumn,showpacs,preprintnumbers,amssymb,superscriptaddress,10pt]{revtex4-1}
\pdfoutput=1

\usepackage{amsmath,amssymb,amsfonts,graphicx,epsfig,todonotes}
\usepackage[noconfig]{refstyle}
\usepackage[hyperfootnotes=true, linktocpage=true, citecolor=blue, linkcolor=blue, urlcolor=Maroon, filecolor=Maroon]{hyperref}

\let\eqref=\relax
\newref{eq}{name={},Name={Eq.~},names={eqs.~},Names={Eqs.~},rngtxt={-},refcmd=(\ref{#1})}
\newref{tab}{name={Table~},Name={Table~},names={tables~},Names={Tables~}}
\newref{sec}{name={Section~},Name={Section~},names={sections~},Names={Sections~}}
\newref{fig}{name={Figure~},Name={Figure~},names={figures~},Names={Figures~}}
\numberwithin{equation}{section}

\newcommand{\IP}{\mathbb{P}}

\newcommand{\comment}[1]{}

\begin{document}

\title{Machine Learning Calabi-Yau Four-folds}

\author{Yang-Hui He}
\affiliation{Department of Mathematics, City, University of London, London EC1V0HB, UK}
\affiliation{Merton College, University of Oxford, OX1 4JD, UK}
\affiliation{School of Physics, NanKai University, Tianjin, 300071, P.R. China}
\email[]{hey@maths.ox.ac.uk}

\author{Andre Lukas}
\affiliation{Rudolf Peierls Centre for Theoretical Physics, University of Oxford, \\Parks Road, Oxford OX1 3PU, UK}
\email[]{andre.lukas@physics.ox.ac.uk}


\begin{abstract}\noindent
Hodge numbers of Calabi-Yau manifolds depend non-trivially on the underlying manifold data and they present an interesting challenge for machine learning. In this letter we consider the data set of complete intersection Calabi-Yau four-folds, a set of about $900,000$ topological types, and study supervised learning of the Hodge numbers $h^{1,1}$ and $h^{3,1}$ for these manifolds. We find that $h^{1,1}$ can be successfully learned (to 96\% precision) by fully connected classifier and regressor networks. While both types of networks fail for $h^{3,1}$, we show that a more complicated two-branch network, combined with feature enhancement, can act as an efficient regressor (to 98\% precision) for $h^{3,1}$, at least for a subset of the data. This hints at the existence of an, as yet unknown, formula for Hodge numbers.
\end{abstract}




\maketitle

\section{Introduction}
\noindent Topological quantities of manifolds, such as Betti or Hodge numbers, are often non-trivially related to the data describing the underlying manifold and tend to be difficult to work out. Explicit formulae are usually not known and calculations rely on complicated and frequently computationally intense algorithms (see, for example, the volume \cite{comp} and references therein for applications of computational algebraic geometry to string and gauge theories). For this reason, such topological properties are an interesting and challenging playground for machine learning. At the most basic level, we can ask if neural  networks are capable of learning these properties. In this letter, we will address this problem for complete intersection Calabi-Yau (CICY) four-folds  and their Hodge numbers.

The complete set of CICY three-folds was the first large dataset of Calabi-Yau manifolds to be constructed~ \cite{Candelas:1987kf,Green:1987cr}. It consists of 7890 different topological types of manifolds which have provided string theoriests and mathematicians alike with a fertile ground for exploration (for some recent applications in the context of string theory, see, for example, Refs.~\cite{Braun:2010vc,Lukas:2017vqp,Anderson:2017aux,Anderson:2007nc,Anderson:2013xka}). More recently, techniques of machine learning have been applied to the study of the string landscape \cite{He:2017aed,He:2017set,Krefl:2017yox,Ruehle:2017mzq,Carifio:2017bov,Deen:2020dlf} (for reviews see Refs.~\cite{He:2018jtw,Ruehle:2020jrk}). In fact, CICY three-folds were the first data set to be analysed from this viewpoint~\cite{He:2017aed}. Subsequent work has studied Hodge numbers of CICY three-folds systematically, using different types of neural network architectures~\cite{Bull:2018uow,Bull:2019cij,Erbin:2020srm,Erbin:2020tks}. 

With the advent of F-theory, Calabi-Yau four-folds have become increasingly important for string compactifications. CICY four-folds have been classified more recently~\cite{Gray:2013mja} and their relevant topological properties have been computed in Ref.~\cite{Gray:2014fla}. The dataset is considerably larger and richer than the one for CICY three-folds and it consists of about $900000$ topological types of manifolds. However, so far, this new dataset has not been used for machine learning and the purpose of this letter is to fill this gap. More specifically, we will explore, within the context of supervised learning, if and to what extent Hodge numbers of CICY four-folds can be learned by neural networks.


\section{Background and notation}\seclabel{s:cicy4}
\subsection{CICY four-folds}
\noindent A CICY four-fold is defined as a complete intersection of the zero loci of $K$ multi-homogeneous polynomials in the ambient space $A = \IP^{n_1} \times \IP^{n_2} \times \ldots \times  \IP^{n_m}$ with dimension $d=n_1+\cdots +n_m=K+4$. The degrees of these polynomials are collected in a $m\times K$ {\em configuration matrix}  $Q=({q^i}_a)$, where $i=1,\ldots ,m$ and $a=1,\ldots ,K$. Its entries ${q^i}_a\in\mathbb{Z}^{\geq 0}$ specify the degree of homogeneity of the $a^{\rm th}$ defining polynomial in the homogeneous coordinates of the $i^{\rm th}$ projective ambient space factor. The Calabi-Yau condition 
\begin{equation}
 n_i+1=\sum_{a=1}^K{q^i}_a\quad\mbox{for}\quad i=1,\ldots ,m\; ,
\end{equation} 
fixes the dimensions $n_i$ of the projective spaces in terms of the configuration matrix. A configuration matrix $Q$, therefore, determines the ambient space $A$ as well as a family of CICY four-folds therein, defined by all (sufficiently generic) polynomials with the specified multi-degrees. Fortunately, many topological quantities, including Hodge numbers, only depend on the family and, hence, only the configuration matrix $Q$, rather than the specific choice of polynomials. 

Calabi-Yau three-folds have two non-trivial~ \footnote{Throughout this letter, we only consider smooth, compact Calabi-Yau $n$-folds with holonomy group $SU(n)$.}
Hodge numbers, $h^{1,1}$ and $h^{2,1}$, which are related to the Euler number $\chi$ by $\chi=2(h^{1,1}-h^{2,1})$. Since the Euler number is usually easily determined, three-folds require only one non-trivial Hodge number computation. The situation is significantly more complicated for four-folds which have four non-trivial Hodge numbers, $h^{1,1}$, $h^{2,1}$, $h^{3,1}$ and $h^{2,2}$. In addition to the formula for the Euler number 
\begin{equation}
\chi = 6 ( 8 + h^{1,1} + h^{3,1} - h^{2,1}) \ ,
\end{equation}
there is an additional linear relation \cite{Sethi:1996es}
\begin{equation}
h^{2,2} = 2 (22 + 2h^{1,1} + 2 h^{3,1} - h^{2,1} ) \label{linrel}\ .
\end{equation}
between those Hodge numbers which can be derived from the index theorem. As for three-folds, the Euler number is usually easily computed. In fact, for CICYs of any dimension it can be expressed explicitly in terms of the entries of the configuration matrix $Q$ (see, for example, Ref.~\cite{huebsch}). In view of Eq.~\ref{linrel}, this leaves us with two Hodge numbers to be determined  by a non-trivial computation and, for our purposes, we will take these to be $h^{1,1}$ and $h^{3,1}$.

CICY four-folds for which the entire second cohomology descends from the ambient space are called {\em favourable} and a significant fraction of CICY four-folds have this property. Evidently, favourable four-folds satisfy $h^{1,1}=m$ so in this case one of the remaining Hodge number computations is simple.

\subsection{Data sets}

\begin{figure}[h!!!]
\[
\begin{array}{c}
\begin{array}{l}\includegraphics[width=3.2in,angle=0,trim= 10mm 0mm 0mm 0mm,clip]{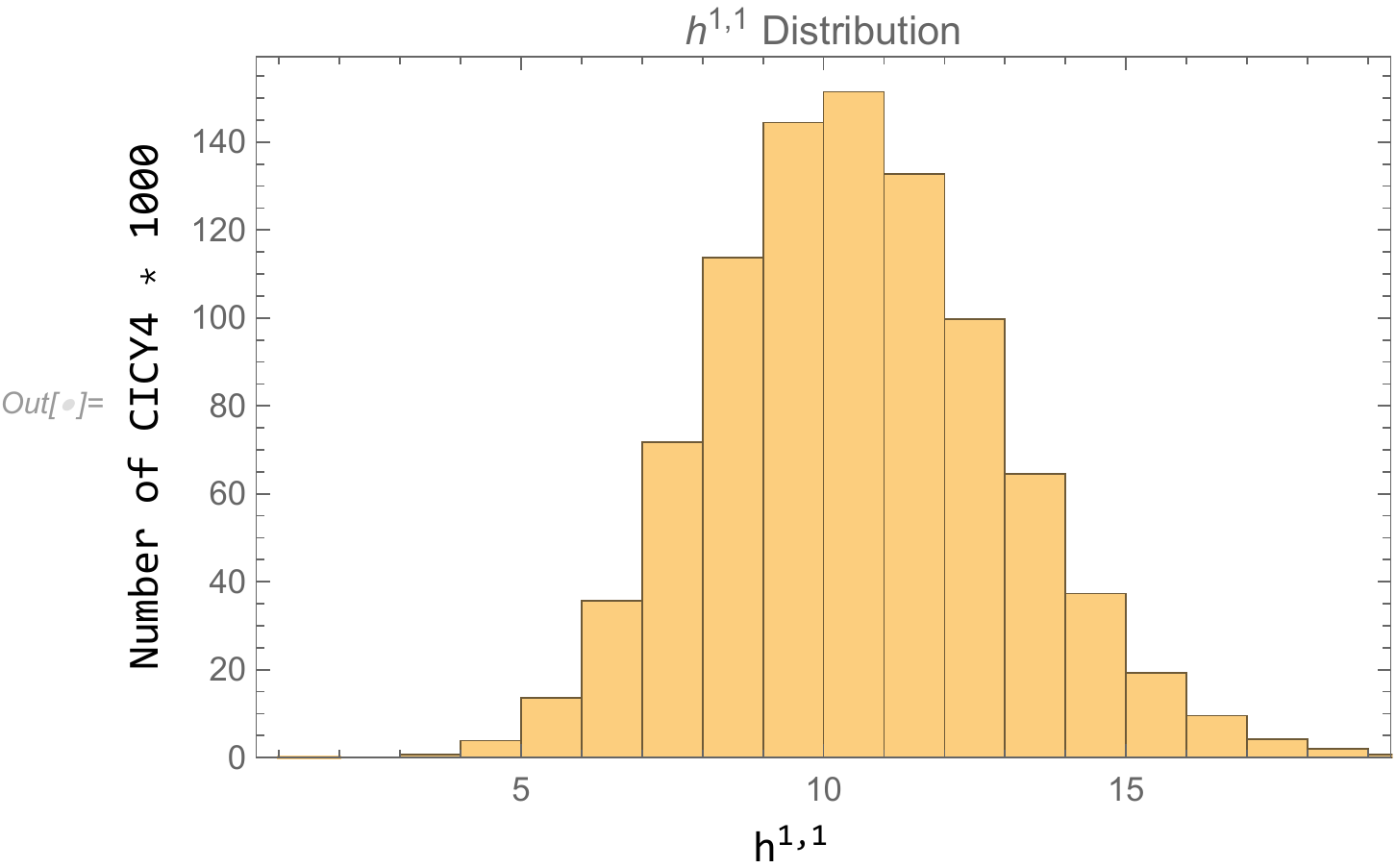}\end{array}
\\
\begin{array}{l}\includegraphics[width=3.2in,angle=0,trim= 10mm 0mm 0mm 0mm,clip]{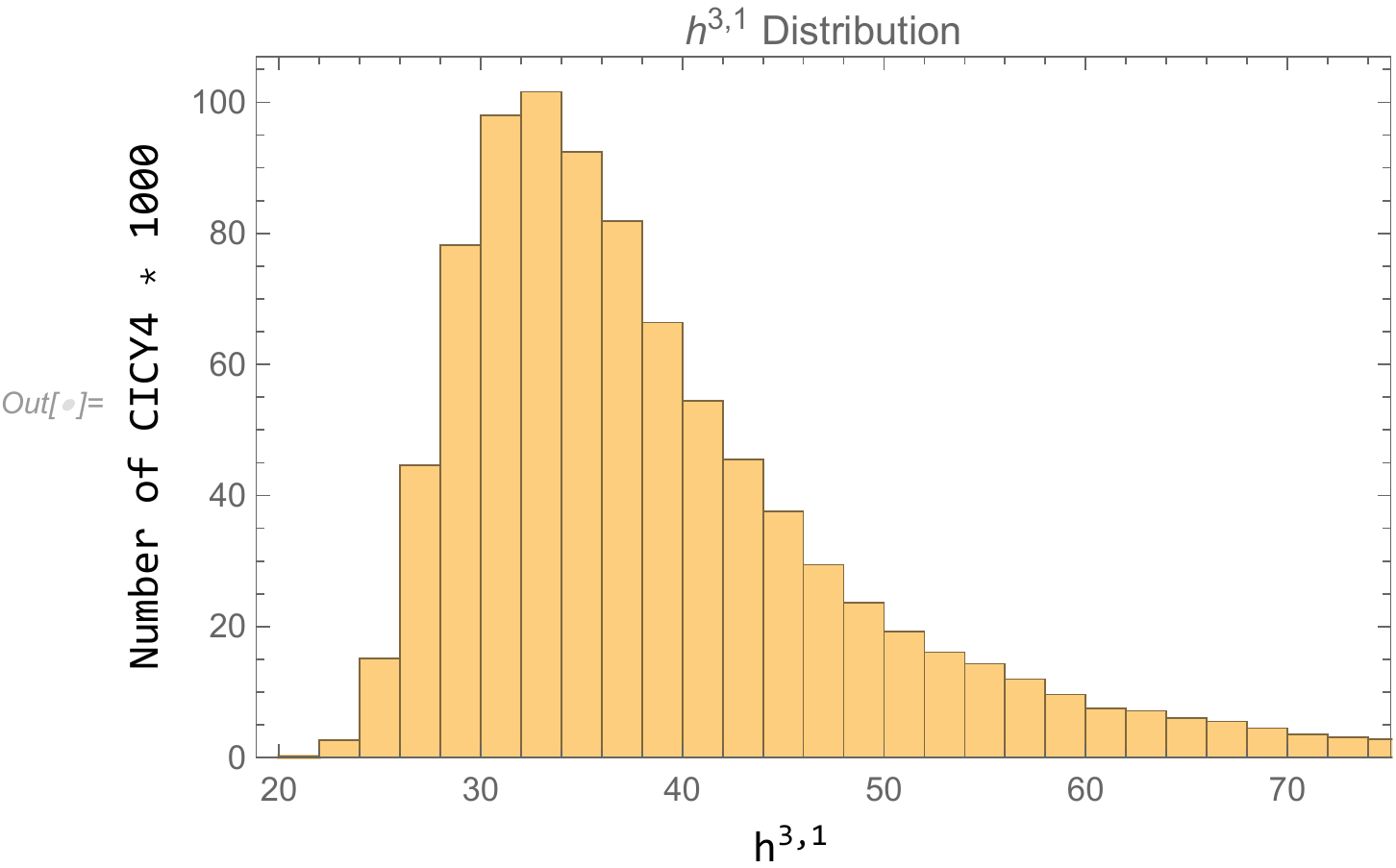}\end{array}
\end{array}
\]
\caption{Distribution of $h^{1,1}$ and $h^{3,1}$ for all CICY four-folds.}
\label{f:histo}
\end{figure}
\begin{figure}[h!!!]
\begin{center}
$
\begin{array}{l}\includegraphics[width=8.5cm]{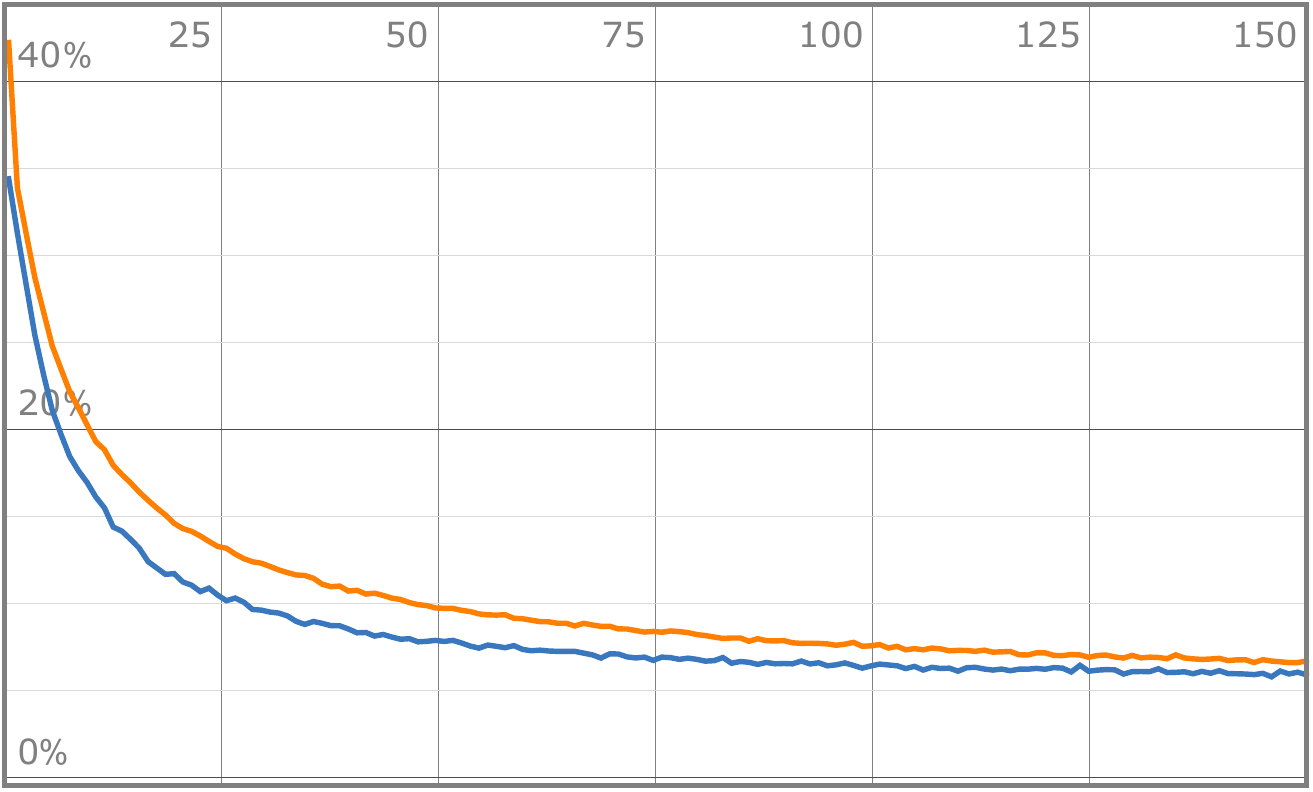}\end{array}
$
\caption{The training plot for the data set $D^{1,1}$ in Eq.~\ref{dataset} and the classifier network~\ref{h11classNN}. Indicated is the error rate as a function of training rounds for the training set (orange) and the validation set (blue).}
\end{center}
\label{f:classifierNN}
\end{figure}
\noindent The different topological types of CICY four-folds were classified in Ref.~\cite{Gray:2013mja} (q.v.~\cite{Gray:2014fla,Kimura:2016crs}) by listing their configuration matrices. Discarding cases which correspond to direct product manifolds, this has led to $905684$ inequivalent configuration matrices $Q$ with minimal size $(m,K)=(1,1)$ and maximal size $(m,K)=(16,20)$. About 54\% of these are favourable. The distribution of Hodge numbers $h^{1,1}$ and $h^{3,1}$ for this data set is shown in Fig.~\ref{f:histo}.

The configuration matrices have different sizes so, as stands, they are not well-suited for training neural networks. We resolve this problem by padding each configuration $Q$ with zeros (on the right and at the bottom) to create a $16\times 20$ enhanced configuration matrix $\tilde{Q}$, whose size matches that of the largest configuration.
As an example, the enhanced configuration matrix $\tilde{Q}$ for a configuration $Q$ with $(m,K)=(10,12)$ is given by
\begin{equation}\nonumber
\tilde{Q}={\tiny
\left(
\arraycolsep=0.1pt\def\arraystretch{0.5}
\begin{array}{cccccccccccccccccccc}
 0 & 0 & 0 & 0 & 0 & 0 & 0 & 0 & 0 & 1 & 1 & 0 & 0 & 0 & 0 & 0 & 0 & 0 & 0 & 0 \\
 0 & 0 & 0 & 0 & 0 & 0 & 0 & 1 & 1 & 0 & 0 & 0 & 0 & 0 & 0 & 0 & 0 & 0 & 0 & 0 \\
 0 & 0 & 0 & 0 & 0 & 0 & 1 & 0 & 1 & 0 & 0 & 0 & 0 & 0 & 0 & 0 & 0 & 0 & 0 & 0 \\
 0 & 0 & 0 & 0 & 1 & 1 & 0 & 0 & 0 & 0 & 0 & 0 & 0 & 0 & 0 & 0 & 0 & 0 & 0 & 0 \\
 0 & 0 & 0 & 1 & 0 & 1 & 0 & 0 & 0 & 0 & 0 & 0 & 0 & 0 & 0 & 0 & 0 & 0 & 0 & 0 \\
 0 & 0 & 2 & 0 & 0 & 0 & 0 & 0 & 0 & 0 & 0 & 0 & 0 & 0 & 0 & 0 & 0 & 0 & 0 & 0 \\
 0 & 1 & 0 & 0 & 0 & 1 & 0 & 1 & 0 & 0 & 0 & 0 & 0 & 0 & 0 & 0 & 0 & 0 & 0 & 0 \\
 1 & 0 & 0 & 0 & 1 & 0 & 0 & 0 & 1 & 0 & 0 & 0 & 0 & 0 & 0 & 0 & 0 & 0 & 0 & 0 \\
 1 & 0 & 0 & 1 & 0 & 0 & 0 & 0 & 0 & 0 & 1 & 0 & 0 & 0 & 0 & 0 & 0 & 0 & 0 & 0 \\
 0 & 1 & 1 & 0 & 0 & 0 & 1 & 0 & 0 & 1 & 0 & 0 & 0 & 0 & 0 & 0 & 0 & 0 & 0 & 0 \\
 0 & 0 & 0 & 0 & 0 & 0 & 0 & 0 & 0 & 0 & 0 & 0 & 0 & 0 & 0 & 0 & 0 & 0 & 0 & 0 \\
 0 & 0 & 0 & 0 & 0 & 0 & 0 & 0 & 0 & 0 & 0 & 0 & 0 & 0 & 0 & 0 & 0 & 0 & 0 & 0 \\
 0 & 0 & 0 & 0 & 0 & 0 & 0 & 0 & 0 & 0 & 0 & 0 & 0 & 0 & 0 & 0 & 0 & 0 & 0 & 0 \\
 0 & 0 & 0 & 0 & 0 & 0 & 0 & 0 & 0 & 0 & 0 & 0 & 0 & 0 & 0 & 0 & 0 & 0 & 0 & 0 \\
 0 & 0 & 0 & 0 & 0 & 0 & 0 & 0 & 0 & 0 & 0 & 0 & 0 & 0 & 0 & 0 & 0 & 0 & 0 & 0 \\
 0 & 0 & 0 & 0 & 0 & 0 & 0 & 0 & 0 & 0 & 0 & 0 & 0 & 0 & 0 & 0 & 0 & 0 & 0 & 0 \\
\end{array}
\right)} 
\quad\leftrightarrow\quad
\begin{array}{l}\includegraphics[width=1in,angle=0]{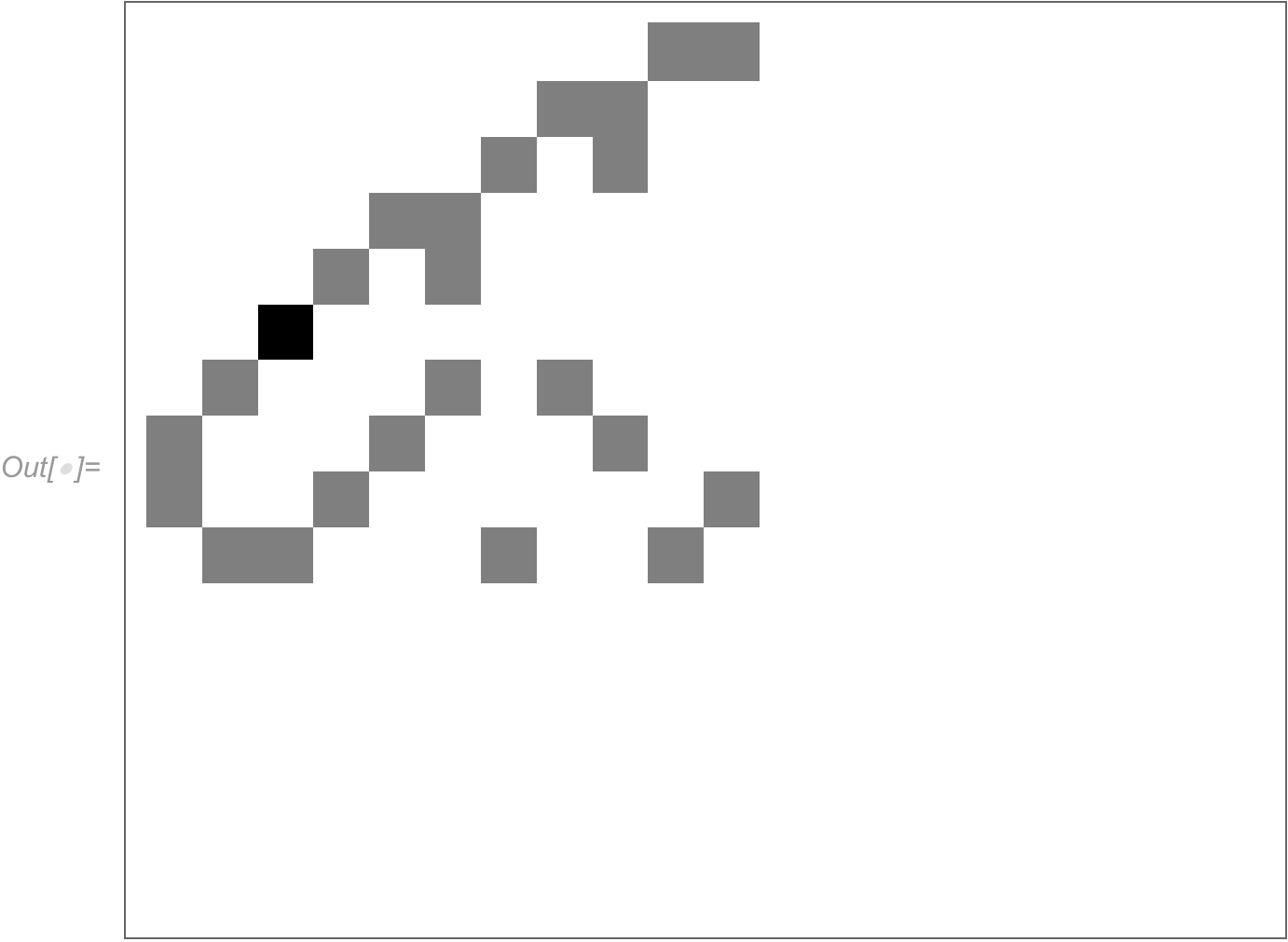}\end{array}
\label{config}
\end{equation}
On the right, we have represented $\tilde{Q}$ by an image with $16\times 20$ pixels and the typical entries $0,1,2$ of $\tilde{Q}$ mapped to grey-scales.
This has been done, as was in Ref.~\cite{He:2017aed}, to emphasise the analogy of our problem with pattern recognition. However, unlike for typical pattern recognition problems (such as classifying the MNIST numbers), it is not intuitively clear how our target (the Hodge numbers) are related to the features of the image.

Our data sets will be of the form
\begin{equation}
 D^{1,1}=\{\tilde{Q}\rightarrow h^{1,1}\}\quad\mbox{or}\quad
 D^{3,1}=\{\tilde{Q}\rightarrow h^{3,1}\}\; . \label{dataset}
\end{equation}
It is possible to enlarge these data sets by adding equivalent configurations obtained from the given ones by simultaneous permutations of rows and columns. Indeed, this method has been used to enlarge the set of CICY three-folds in Refs.~\cite{He:2017aed,He:2017set,Bull:2018uow,Bull:2019cij}. However, the set of CICY four-folds is considerably larger and numbers are certainly sufficient for machine learning purposes without any enlargement. In fact, for cases where we use the entire data set we have also checked that enlarging by equivalent configurations does not significantly increase the performance of the networks we study. For these reasons, we take the data sets $D^{1,1}$ and $D^{3,1}$ above to contain the 905684 inequivalent (enhanced) configurations of the original classification.

We will also analyse {\em feature-enhanced} versions of these data sets where we supplement the configuration matrix $Q$ by monomials of degree up to four in its entries ${q^i}_a$. For larger configuration matrices this leads to very large input spaces which are not practical. For this reason and for specificity we will limit our discussion to configurations with size $(m, K)=(4,4)$. This subset only contains $1035$ configurations so, unlike for the full data set, we now opt for an enlargement by all simultaneous row and column permutations of $Q$. This leads to a total of  $1035\times 24^2=596160$ configurations. These $4\times 4$ configurations $Q$ are then feature-enhanced to a vector $Q_q$, where $q=1,2,3,4$, by including all monomials of degree $\leq q$ between the column entries of $Q$. For $q=2$ this means $Q_2=({q^i}_a,{q^j}_a{q^k}_a)$, where $a,i,j,k=1,\ldots ,4$ and $j\leq k$, and analogously for $q>2$. The dimensions 
\footnote{Recall that one can write ${m+d-1 \choose d}$ independent monomials of degree $d$ in $m$ variables (here $m=4$), so there are $10$ quadrics, $20$ cubics, and $35$ quartics composed from the columns of $Q$.} 
$d_q$ of these enhanced configurations are given by $(d_1,d_2,d_3,d_4)=(4\times 4,14\times 4,34\times 4,69\times 4)$. In summary, this leads to data sets of the form
\begin{equation}
 D^{1,1}_q=\{Q_q\rightarrow h^{1,1}\}\quad\mbox{or}\quad
 D^{3,1}_q=\{Q_q\rightarrow h^{3,1}\}\; , 
 \label{datasetsenh}
\end{equation}
where $q=1,2,3,4$. As is customary, we need to {\it disjointly} split the above data sets into a training set, a validation set and a test set. We typically use 15\% for training and 5\% for validation, both randomly selected from the full set, and the remainder of 80\% for testing. The validation set is used to monitor progress during training and we evaluate the trained network on the test set.

\subsection{Neural Networks}
\noindent Key components in the subsequent discussion are standard forward-feed, fully connected neural networks of depth $d$, which define a map of the form
\begin{equation}\nonumber
 \mathbb{R}^{n_0}\stackrel{L_{n_1}}{\longrightarrow}\mathbb{R}^{n_1}\stackrel{f}{\longrightarrow}\mathbb{R}^{n_1}\stackrel{L_{n_2}}{\longrightarrow}\cdots\stackrel{L_{n_d}}{\longrightarrow}\mathbb{R}^{n_d}\stackrel{f}{\longrightarrow}\mathbb{R}^{n_d}\; .
\end{equation} 
Here $L_n$ is a standard affine transformation with trainable weights and biases and co-domain dimension $n$ and $f$ represents a component-wise function, typically a logistic sigmoid function,  $\sigma(z) := (1 + e^{-z})^{-1}$, or a scaled exponential-linear unit (SELU), defined by $s(z)=1.0507\, z$ for $z\geq 0$ and $s(z)=1.7851(\exp(z)-1)$ for $z<0$. In some cases, we will use a probability $p$ dropout layer, denoted $\delta_p$, which is a standard tool to avoid over-fitting. The dropout probability $p$ is chosen to optimise performance. For classifier networks we also require a softmax layer $S(z_i) = e^{z_i} (\sum_i e^{z_i})^{-1}$.  For notational convenience, we will use the short-hand ${\cal N}_{n_0}(n_1,f,n_2,\ldots ,n_d,f)$ for the above network.

Explicit training is carried out with the Mathematica machine learning suite \cite{wolfram}, using the ADAM \cite{ADAM} steepest gradient descent minimiser and a mean square loss. 
Evidently, the network architectures explored in this letter are relatively simple. We have checked that convolutional networks, similar to those used for digit recognition, do not improve the performance significantly. However, it would be expedient to apply the methods of Ref.~\cite{Erbin:2020srm,Erbin:2020tks}, as well as the interesting representation of configurations in \cite{Krippendorf:2020gny} to the CICY four-fold data set.

\section{Learning $h^{1,1}$}\seclabel{h11}

\noindent Figure~\ref{f:histo} shows that $h^{1,1}$ takes a rather limited set of values for our data set. More specifically, it turns out that $h^{1,1}\in\{1,2,\ldots ,24\}$. This suggest that both a $24$-way classifier network and a regressor network with a real output intended as an approximation of $h^{1,1}$ may be feasible. We discuss these two options in turn.
\subsection{Classifier network}
\noindent The relevant data set for this task is $D^{1,1}$ in Eq.~\ref{dataset} which is used to train a network of the form
\begin{equation}
 N_{16\times 20}(512,\sigma,\delta_{0.4},256,\sigma,\delta_{0.3},256,\sigma,24,S)\; . \label{h11classNN}
\end{equation}
As mentioned earlier, we use 15\% of the data set for training and 5\% for validation. Training is performed at a learning rate of $1/300$ for $150$ rounds and takes about $18$ minutes on a single laptop CPU. The training curves are shown in Fig.~\ref{f:classifierNN}.
The trained network is applied to the test set (at 80\% the bulk of the data) and it predicts $h^{1,1}$ correctly for 96\% of the cases. The $24\times 24$ confusion matrix is diagonal to a good accuracy, with any single off-diagonal entry $<0.05$.
This is a rather convincing performance by a relatively simple, feed-forward network. We note that the substantial width of the network~\ref{h11classNN} is required to achieve the stated accuracy and we have to include the dropout layers in order to avoid over-fitting. In summary, we conclude that the Hodge numbers $h^{1,1}$ for CICY four-folds can be successfully learned by a suitably configured fully connected classifier network. 

Not surprisingly, for favourable manifolds, the network predicts $h^{1,1}$ with 100\% accuracy, so misclassifications only arise for non-favourable cases. This observation suggest that a simple binary classifier network, similar to \ref{h11classNN} but with the $24$-dimensional output layer replaced by a two-dimensional one, can be used to distinguish favourable and non-favourable CICY four-folds. This is indeed the case and works at about 96\% accuracy on the test set.  

The above network generalises well when trained on a randomly selected training set. A somewhat more ambitious question is whether a network trained on configurations with small Hodge number, say $h^{1,1} < 8$ (about 20\% of the configurations), can predict the Hodge numbers of configurations with $h^{1,1}\geq 8$. For CICY three-folds this was attempted in Ref.~\cite{Bull:2019cij}. Obviously, such a network, trained only on small and relatively simple configurations but able to predict properties of larger and more complicated ones would be very useful. Unfortunately, for the case of CICY four-folds and classifier networks of the type~\ref{h11classNN} this does not work well and the network performs poorly, with a success rate close to zero on configurations with $h^{1,1}\geq 8$. However, seeding the training set with a small sample (say 10000) of configurations with $h^{1,1}\geq 8$ leads to a significant improvement (success rate around 0.6).
\subsection{Regressor network}
\noindent Encouraged by the success of the classifier, let us see how a regressor performs. 
We use the same dataset $D^{1,1}$ in Eq.~\ref{dataset} and a network of the form
\begin{equation}
 {\cal N}_{16\times 20}(512,s,256,s,128,s,32,s,8,s,1)\; . \label{h11regNN}
\end{equation} 
The idea is that the one-dimensional real output of this network approximates $h^{1,1}$ and we take its rounding to the nearest integer as the prediction for $h^{1,1}$. This is clearly challenging since a successful prediction requires an accurately trained network with a typical loss significantly less than one.

The above network is the best-performing we have found. After training for 150 rounds at a learning rate of 1/1000 (about 15 minutes on a single CPU), the network output has an average deviation from $h^{1,1}$ of $\sim 0.3$ on the test set.
This translates, after rounding, into 83\% of test set values correctly predicted. While this is a respectable success rate and the network trains efficiently, a wrong prediction for $h^{1,1}$, typically by $1$, in 17\% of the cases means the network is of limited practical use.

\section{Learning $h^{3,1}$}
\noindent  Fig.~\ref{f:histo} shows that the range of $h^{3,1}$ values is considerably larger than the one for $h^{1,1}$. More specifically, we have $20\leq h^{3,1}\leq 426$. As we will see, for this range it is significantly harder to obtain convincing performances from simple classifier or regressor networks of the kind we have used for $h^{1,1}$. For this reason, we also explore other options, focusing on the feature-enhanced data sets $D_q^{3,1}$ in Eq.~\ref{datasetsenh} and more complicated two-branch networks.
\subsection{Classifier and regressor networks}
\noindent A $407$-way classifier based on a network of the form
\begin{equation}\nonumber
 N_{16\times 20}(512,\sigma,\delta_{0.4},512,\sigma,\delta_{0.4},512,\sigma,\delta_{0.4},407,S)
\end{equation}
trained on the dataset $D^{3,1}$ in Eq.~\ref{dataset} leads to a poor performance, with a 27\%  success rate on the test set. Likewise, a regressor network of the form
\begin{equation}\nonumber
 N_{16\times 20}(256,s,\delta_{0.2},128,s,16,s,1)\; ,
\end{equation}
trained on $D^{3,1}$, produces test set predications for $h^{3,1}$ with an average deviation of $\sim 2.7$ from the true value. While this might be considered a reasonable accuracy for some purposes, it is not sufficient to predict the correct integer after rounding. In fact, only 15\% of test set values for $h^{3,1}$ are reproduced exactly after rounding. 
For either of the above networks, we have not been able to improve performance significantly by hyper-parameter optimisation.

\subsection{Classifier and regressor for $4\times 4$ configurations}
\noindent We can ask if a classifier network performs better on the data set $D_1^{3,1}$ of $4\times 4$ configurations as defined in Eq.~\ref{datasetsenh}. In addition to a much smaller dimension of the feature space,  the range of $h^{3,1}$ values is now reduced to $20\leq h^{3,1}\leq 260$. In fact, a $235$-way classifier network of the form
\begin{equation}\nonumber
{\cal N}_{4\times 4}(512,\sigma,\delta_{0.4},512,\sigma,\delta_{0.4},512,\sigma,\delta_{0.4},235,S)\; ,
\end{equation}
trained on $D_1^{3,1}$ performs perfectly on the test set at a 100\% success rate.  This is quite impressive, considering the number of classes is still large.

However, a regressor network of the form
\begin{equation}\nonumber
 {\cal N}_{4\times 4}(512,s,256,s,64,s,16,s,1)
\end{equation}
trained on $D_1^{3,1}$ is much less successful. It predicts $h^{3,1}$ for the test set with an average error of $\sim 1$ which leads to a success rate of 35\% after rounding. We have not been able to improve this performance significantly by variations in hyper-parameters.

\subsection{Two branch network and feature enhancement} 
\noindent Is it possible to construct a successful regressor network for $h^{3,1}$? The approach we are about to present is motivated by observations made in the related context of line bundle cohomology. Line bundle cohomology dimensions have been conjectured, in many cases empirically verified~\cite{Constantin:2018hvl,Klaewer:2018sfl,Larfors:2019sie} and for some classes shown~\cite{Brodie:2019pnz} to be described by piecewise polynomial formulae in the line bundle degrees. The degree of the polynomials equals the complex dimension of the underlying manifold. In Ref.~\cite{Brodie:2019dfx} a two-branch neural network adapted to this structure and trained with feature-enhanced data has been constructed. It has been shown that conjectures for piecewise polynomial cohomology formulae can be extracted from this network. 

The present context is of course somewhat different. We are not interested in all line bundles on a fixed manifold but rather in specific properties for a class of different manifolds. Nevertheless, it is the case that computations of Hodge numbers for CICYs are ultimately reduced to the computation of line bundle cohomology. For this reason it is not far-fetched to try a two-branch regressor network, similar to the one used in Ref.~\cite{Brodie:2019dfx}, in order to learn $h^{3,1}$.

More specifically, we would like to consider networks of the form
\begin{equation}\nonumber
\begin{array}{ccrcc}
Q_1&\rightarrow&{\cal N}_{4\times 4}(512,\sigma,512,\sigma,256,\sigma)&&\\[-2mm]
&&&\searrow&\\
&&&&\mbox{dot}\;\rightarrow\; h^{3,1}\\
&&&\nearrow&\\[-2mm]
&&Q_q\;\rightarrow\;{\cal N}_{d_q}(256)
\end{array}
\end{equation}
where ``dot" indicates a dot product between the two vectors. The upper branch of the network is intended to detect the regions of the underlying piecewise polynomial formula. Since the boundaries of these regions are usually described by linear equations the upper branch only receive the $4\times 4$ configuration matrices $Q_1\in D_1^{3,1}$. On the other hand, the lower part of the network, which consists of a single affine layer is supposed to reproduce the polynomial and, therefore, receives the feature-enhanced matrices $Q_q\in D_q^{3,1}$ which consists of the configuration matrix as well as its monomials of degree $\leq q$. The analogy with line bundles suggests that we need up to quartic monomials (since we are working on four-folds). We have, therefore, constructed the data sets $D^{3,1}_q$ for $q$ up to four. For comparison purposes we will consider all cases $q=1,2,3,4$.

Training the above network with $D^{3,1}_1$ and $D^{3,1}_q$ for $q=1,2$ leads to poor performance, with an average error of $\sim 1$ and a test set success rate of 54\% for $q=1$ and 40\% for $q=2$.
On the other hand, training with $D^{3,1}_1$ and $D^{3,1}_q$ for $q=3,4$ leads to very accurately trained networks. Specifically, for $q=3$ we achieve an average error of $0.04$ which translates into a 98\% success rate on the test set. For $q=4$ the results are similar, with an average error of $0.17$ and a test set success rate of 95\%.

Achieving this accuracy for $q=3,4$ requires a careful adjustment of the learning rate during training. In an initial training step of about $100$ rounds the learning rate is set to $1/1000$. This leads to a network whose average error does not decrease below $\sim 1$. Adding successive short training steps of about 5 - 10 rounds with gradually decreasing learning rate to a final value of $1/100000$ then leads to the accuracy mentioned above. 

In conclusion, we are able to build a successful regressor for $h^{3,1}$, at least for the $4\times 4$ configurations under consideration, by using a two branch network motivated by the results for line bundle cohomology in Ref.~\cite{Brodie:2019dfx}. As expected, we require feature-enhanced data which includes at least quadrics and cubics of the configuration matrix for this network to perform well.
We note that the two-branch network can also be applied to the data sets $D_q^{1,1}$ for $h^{1,1}$, where it leads to a 100\% success rate on the test set.

\section{Conclusion \& Outlook}\seclabel{conclusion}
\noindent Computing Hodge numbers of Calabi-Yau manifolds is a non-trivial task and presently known methods require, in all but special cases, complicated algorithms in commutative algebra, based on sequence-chasing in cohomology. 
For this reason, machine learning of Hodge numbers is an interesting and challenging task. This problem has obvious analogies with image classification, as originally  pointed out in Ref.~\cite{He:2017aed}. Despite this analogy, it is, a priori, unclear  if these numbers can be successfully learned and, if so, what the required network architectures might be. 

In this letter, we have studied supervised machine learning of the Hodge numbers $h^{1,1}$ and $h^{3,1}$ for complete intersection Calabi-Yau (CICY) four-folds. This data set consists of about 900,000 topological types, each described by an integer (configuration) matrix $Q$. We find that $h^{1,1}$ can be successfully predicted from $Q$ with both fully connected classifier and regressor networks. The former are particularly effective and lead to a 96\% success rate on the test set when trained on only 15\% of the data.

Unfortunately, fully connected classifier or regressor networks do not work efficiently for $h^{3,1}$, presumably due to the large range of $h^{3,1}$ values. However, we have shown that a two branch regressor network, combined with feature enhanced data, works well for a subset of the data which consists of $4\times 4$ configuration matrices.

The structure of this two-branch network is motivated by recent results for line bundle cohomology~\cite{Constantin:2018hvl,Klaewer:2018sfl,Larfors:2019sie,Brodie:2019pnz}. Its success hints at the existence of a formula for $h^{3,1}$ in terms of $Q$ which is at present unknown. It would be interesting to search for this formula, possibly assisted by the information encoded in the trained network. Such a formula would be a new mathematical result and useful for applications in theoretical physics.

\vskip 2mm
\noindent
{\bf Acknowledgments} YHH would like to thanks STFC UK, for grant ST/J00037X/1. AL would like to thank Andrei Constantin for discussions.


\end{document}